\documentclass[twocolumn]{webofc}

\usepackage[varg]{txfonts}   
\usepackage{amsmath}
\usepackage{hyperref}
\usepackage{cleveref}

\hypersetup{
  pdftitle={A Study of Modified Characteristics of Hadronic Interactions},
  pdfauthor={Jiri Blazek},
  pdflang=en
}

\newcommand{\Nmu}{$N_{\mu}$}
\newcommand{\sib}[1]{{\sc Sibyll}~#1\xspace}
\newcommand{\qgsii}{{\sc Qgsjet~II-04}\xspace}
\newcommand{\eposlhc}{{\sc Epos-lhc}\xspace}

\newcommand{\corsika}{{\sc Corsika}\xspace}

\newcommand{\Xmax}{\ensuremath{X_{\rm max}}\xspace}

\newcommand{\Rmu}{\ensuremath{R_{\mu}}\xspace}

\begin{document}
\title{A Study of Modified Characteristics of Hadronic Interactions}
%
%

\author{\firstname{Jiri} \lastname{Blazek}\inst{1}\fnsep\thanks{\email{blazekj@fzu.cz}} \and
        \firstname{Jan} \lastname{Ebr}\inst{1}\fnsep \and
        \firstname{Jakub} \lastname{Vicha}\inst{1}\fnsep \and
        \firstname{Tanguy} \lastname{Pierog}\inst{2}\fnsep \and
        \firstname{Petr} \lastname{Travnicek}\inst{1}\fnsep 
}
\institute{FZU Prague, Czech Republic
\and
Karlsruhe Institute of Technology, Germany
}

\abstract{%
  We have implemented ad-hoc modifications to the CORSIKA Monte-Carlo generator which allow us to simultaneously adjust the multiplicity, elasticity and cross-section of hadronic interactions with respect to the predictions of the Sibyll 2.3d interaction model, in order to assess whether a reasonable combination of changes (that is not excluded by current experimental data) could alleviate the observed tension between the model predictions and observed features of extensive air showers induced by ultra-high energy cosmic rays (UHECR). Previously, we have studied the effects of such changes on proton-initiated showers. Because a multitude of experimental data suggest that the primary composition of the UHECR is mixed, we have expanded the modification procedure to include nuclear projectiles in a consistent way based on the superposition model, in a similar manner as was used in the previous studies carried out using one-dimensional simulation methods. As we are using a fully three-dimensional approach, we can quantify the effects of the changes on both longitudinal and lateral features of the showers. With the inclusion of nuclear projectiles, we can study the impact of the changes on observable quantities for realistic primary beams as well as on the determination of the primary composition from data under the assumption of the modified hadronic interactions.
}

\maketitle
\section{Introduction - The Muon Puzzle}
\label{intro}

The muon puzzle in the physics of ultra-high energy of cosmic-rays (UHECR, energy above $10^{18}$~eV) is a long-standing and intriguing problem that has challenged the current understanding of primary particles and their interactions with the Earth's atmosphere \cite{Albrecht:2021cxw}. In the UHECR region, there exists a well-established tension between the mass composition derived from fluorescence measurements using the maximum of the shower development \Xmax and mass composition inferred from ground-based observational techniques. More specifically, measurements at Earth's surface observe a number of muons \cite{PierreAuger:2016nfk}, \cite{TelescopeArray:2018eph}, \cite{PierreAuger:2014ucz}, \cite{AmigaMuons} which is larger than the predictions of various high-energy hadronic interaction models when considering the composition derived using \Xmax, leading to an interpretation of heavier composition compared to fluorescence techniques. This discrepancy is supported by measurements by various other experiments, and there is evidence that it gets more significant with the rise of energy \cite{Soldin:2021wyv}.

The predictions of hadronic interaction models, such as \sib{2.3d} \cite{Engel:2019dsg}, \qgsii \cite{Ostapchenko:2010vb} and \eposlhc \cite{Pierog:2013ria}, are based on data coming from collider experiments. These measurements of cross sections, hadron production spectra etc. are then extrapolated to the energy and centrality region of ultra-high energy cosmic rays. A thorough overview of the relation of different collider observables to the features of extensive atmospheric air showers is given in \cite{Albrecht:2021cxw}. Naturally, there is a systematic uncertainty associated with the extrapolation. In this work, we will investigate whether modifications applied to features of hadronic interactions - cross section, inelasticity and multiplicity, could possibly lead to the reconcilement of the models' predictions and data. It should be stressed that any such introduced modifications are purely ad-hoc, with no basis in hadronization physics, and should serve only as a guidance to the model builders. We follow the approach introduced by Ulrich et al. in \cite{Ulrich:2010rg} and we build upon our previous result \cite{Blazek:2021Cb} where we implemented a set of conservative modifications for a proton primary. In this work, we introduce also a self-consistent approach to modifying interactions of heavier nuclei and we more thoroughly explore the generated phase space of observable features of extensive air showers such as the \Xmax and the muon signal on ground, \Nmu.

\section{Ad-hoc modifications of hadronic interactions}
\label{modifications}

Following \cite{Ulrich:2010rg}, we consider energy-dependent changes to the number and properties of particles produced by a hadronic interaction generator, in our case \sib{2.3d}, and to the cross-section of the interactions. We utilize this approach, rather than changing the intrinsic parameters of the model, in order to more fully explore the available phase space. The choice of \sib{2.3d}, which uses a superposition model for general nucleon-nucleon interactions, makes this procedure straightforward both on conceptual and technical levels. We make use of the \corsika package, which allows us to obtain also information about the lateral features of an extensive air shower. 

Concretely, we recalculate the following set of three parameters of a hadronic interaction:

\begin{itemize}
\item The hadronic particle cross-section, $\sigma_{\mathrm{prod}}$.
\item The elasticity, defined as $\kappa_{\mathrm{el}} = E_{\mathrm{leading}} / E_{\mathrm{tot}}$, where $E_{\mathrm{leading}}$ is the energy of leading particle in the lab system and  $E_{\mathrm{tot}}$ is the energy of the incoming particle in the interaction. 
\item The secondary multiplicity, defined as the total number of particles escaping the interaction.
\end{itemize}

The modifications are implemented in such a way that obeys all the expected symmetries - most notably the energy, momentum, charge and isospin are conserved. The parameters are modified concurrently. We have already demonstrated in \cite{Blazek:2021Cb} that applying two simultaneous modifications has a small but non-negligible effect when compared with super-imposing the respective changes.

Recent studies have shown that the muon puzzle could be resolved by various mechanisms in which the probabilities to produce particles of different types are affected -- \cite{Baur:2019cpv}, \cite{Anchordoqui:2022fpn}, \cite{Manshanden:2022hgf}. We could have simulated this effect by varying the $\pi^{0} / \pi^{\pm}$ ratio. This would however likely cover our entire phase space, and wouldn't allow us to explore the effect of the other modifications in detail. We thus chose a constrained approach, featuring three modifications with constraints from collider measurements. 

The modification is applied by calculating an energy dependent factor and multiplying the respective parameter.

\begin{equation}
    f(E, f_{19}) = 1 + (f_{19} - 1)\cdot F(E)
\end{equation}

where $F(E) = 0$ below some threshold energy $E_{\mathrm{thr}}$ and otherwise

\begin{equation}
    F(E) = \frac{\mathrm{log}_{10}(E / E_{\mathrm{thr}})}{\mathrm{log}_{10}(\mathrm{10~EeV} / E_{\mathrm{thr}})}.
\end{equation}

The values of $E_{\mathrm{thr}}$ can in principle be different for each modified parameter. The functional form of the scaling was chosen such that it represents the increasing uncertainty of the model's prediction in a straightforward linear fashion. Ideally, we would like to start within $\pm 3~\sigma$ of a known collider measurement at the respective $E_{\mathrm{thr}}$ and extrapolate to UHECR energies. We discuss the choice of the threshold energies more extensively in \cite{Blazek:2021Cb}, here we will just state the outcomes:

\begin{itemize}
    \item For cross section, we set $E_{\mathrm{thr}} = 10^{16}$~eV, corresponding to $f_{19} \subset (0.8, 1.2)$
    \item For multiplicity, we set $E_{\mathrm{thr}} = 10^{15}$~eV, corresponding to $f_{19} \subset (0.6, 1.7)$
    \item For elasticity, we set $E_{\mathrm{thr}} = 10^{14}$~eV, corresponding to $f_{19} \subset (0.6, 1.5)$
\end{itemize}

Note that, especially for the modifications of the cross section, this approach is much more conservative than previous work. 

The technical implementation of the respective modifications is described in detail in \cite{Ulrich:2010rg} and we shall not attempt to reproduce it here. 

The modifications are considered to apply to a proton-air (or, in general, nucleon-air) interaction. For nuclei, the factor $f$ is thus calculated from the energy of the interacting particle per nucleon. The \sib{2.3d}model treats the individual nucleon-air sub-interactions in a nucleus-air collision separately and thus the same procedures for resampling of the produced particles used to adjust the resulting multiplicity and/or inelasticity in proton-air collisions are simply applied to each of the nucleon-air sub-interactions. For cross-section modifications, the precise way of calculating would be to adjust the proton-air cross-section using the factor $f$ (for the appropriate per-nucleon energy), convert it into a modified proton-proton cross-section and then calculate the nucleus-air cross-section from the modified value in the same manner as it is calculated in the model. This is however a complex calculation; moreover, in the implementation of the model in \corsika, the nuclear cross-sections are pre-calculated, making the implementation of this approach cumbersome. To make changes to them on the fly, we use a simpler parametrization based on the observation that in \sib{2.3d}, both proton and nuclear cross-sections at the relevant energies follow very well a simple power law function $\sigma(E)=CE^B$ (with  parameters depending on $A$) and the assumption that the nucleus-air cross-section at energy $E$ is a function of just the proton-air cross-section at $E/A$. Then the modified proton-air cross-section is equal to proton-air cross-section at a different energy

\begin{equation}
\begin{split}
    \sigma_\mathrm{p}^{\mathrm{mod}}(E/A) & = f\sigma_\mathrm{p}(E/A)=fC_\mathrm{p}(E/A)^{B_\mathrm{p}} \\
    & =C_\mathrm{p}(Ef^{-{B_\mathrm{p}}}/A)^{B_\mathrm{p}}=\sigma_\mathrm{p}(Ef^{-B_\mathrm{p}}/A).
\end{split}
\end{equation}

The nucleus-air cross-section corresponding to this proton-air cross section is then simply obtained by evaluating the formula at the corresponding energy
\begin{equation}
\begin{split}
    \sigma_\mathrm{A}^{\mathrm{mod}}(E) & =\sigma_\mathrm{A}(Ef^{-B_\mathrm{p}})=C_\mathrm{A}(Ef^{-B_\mathrm{p}})^{B_\mathrm{A}} \\ & =f^{B_\mathrm{A}/B_\mathrm{p}} \sigma_\mathrm{A}(E);
\end{split}
\end{equation}
so that the factor $f$ is only modified by the ratio of the two exponents. Compared to just using the factor $f$ calculated at $E/A$ as it is, this correction introduces a shift no larger than 0.6 g cm$^{-2}$ in interaction depth for the nuclei, energies and $f_{19}$ values in question, from which we can conclude that any further improvements in this calculation would have a negligible impact on the results.

\section{Simulation setup}

For reasons described in the previous section we utilize the \sib{2.3d} model. We have performed tests with the \qgsii~and \eposlhc~with sparser binning of modification parameters and smaller statistics, using only a proton primary. The choice of a model doesn't have a strong effect  on the general behavior of e.g. the dependence of \Xmax change on $f_{19}$ for the case of a proton primary particle. 

The simulated primary beam consists of protons and iron nuclei of equal energy, 10$^{18.7}$~eV. We simulate the incoming primary particles having two zenith angles, 0 and 60 degrees. Previously, we also simulated intermediate zenith angles with the aim of observing any emerging zenith dependence and eventually constructing a ground based quantity similar to $S_{38}$ utilized by the Pierre Auger Observatory \cite{PierreAuger:2020yab}, \cite{PierreAuger:2020qqz}. Since we did not attempt this approach in this work, we restricted ourselves to only the two extreme zenith angles to save computing time. 

We make use of the \corsika Monte-Carlo generator \cite{Heck:1998vt} in order to capture the lateral features of the shower's development. More precisely, we utilize the option CONEX in CORSIKA, with our modifications that allow for resampling of the hadronic interactions and for steering of the simulation. Particles in the shower cascade are initially treated with the one-dimensional Monte-Carlo CONEX code and every interaction is resampled if its energy is above the respective $E_{\mathrm{thr}}$. The particles are then given over to the CORSIKA code at 300 GeV, which performs a full 3D Monte-Carlo simulation and outputs particle densities at ground level. We set this observational level to 1400 meters above the sea level, in correspondence with the characteristics of the site of the Pierre Auger Observatory. Muons produced within the one-dimensional simulation are handed over to the CORSIKA code immediately at all energies. 

Overall, we carry out simulations for proton and iron primaries, two values of zenith angle, 0 and 60 degrees, with three modifications of the cross-section $f^{\sigma}_{19} = (0.8, 1.0, 1.2)$, five modifications of the elasticity $f^{\mathrm{el}}_{19} = (0.6, 0.8, 1.0, 1.2, 1.5)$ and five modifications in multiplicity $f^{\mathrm{mult}}_{19} = (0.6, 0.8, 1.0, 1.3, 1.7)$, covering a total phase space of 300 combinations. We simulate 1000 showers for every such bin. 

The large generated libraries of 3-dimensional shower simulations allow for the extraction of many observables, related both to the longitudinal development of the showers and to the numbers, types and spectra of particles at the ground. Here we focus on the maximum of the longitudinal development \Xmax (which is straightfowardly extracted from  the longitudinal shower profile saved by CORSIKA) and the number of muons at ground. To extract this number, we sum the number of particles at ground in rings with a given radius, perpendicular to the shower axis. To again facilitate a better comparison with the data from the Pierre Auger Observatory, we set the radius to 1000 meters, which is the distance where Auger reports their results. To maintain independence of any particular experiment, we use simply the number of muons $N_\mu^{1000}$; for a typical detector, the measured signal will be proportional to this number. We track muons down to the threshold energy of 10 MeV. We performed extensive tests in order to verify that the simulations obtained in this hybrid approach are compatible within statistical uncertainties with simulations performed with the full 3D simulation using the standalone CORSIKA treatment, in particular that the shapes of the lateral distribution functions exhibit a good match. 

\section{Results}
\label{sec-2}

\begin{figure}[h]
\centering
\includegraphics[width=8cm,clip]{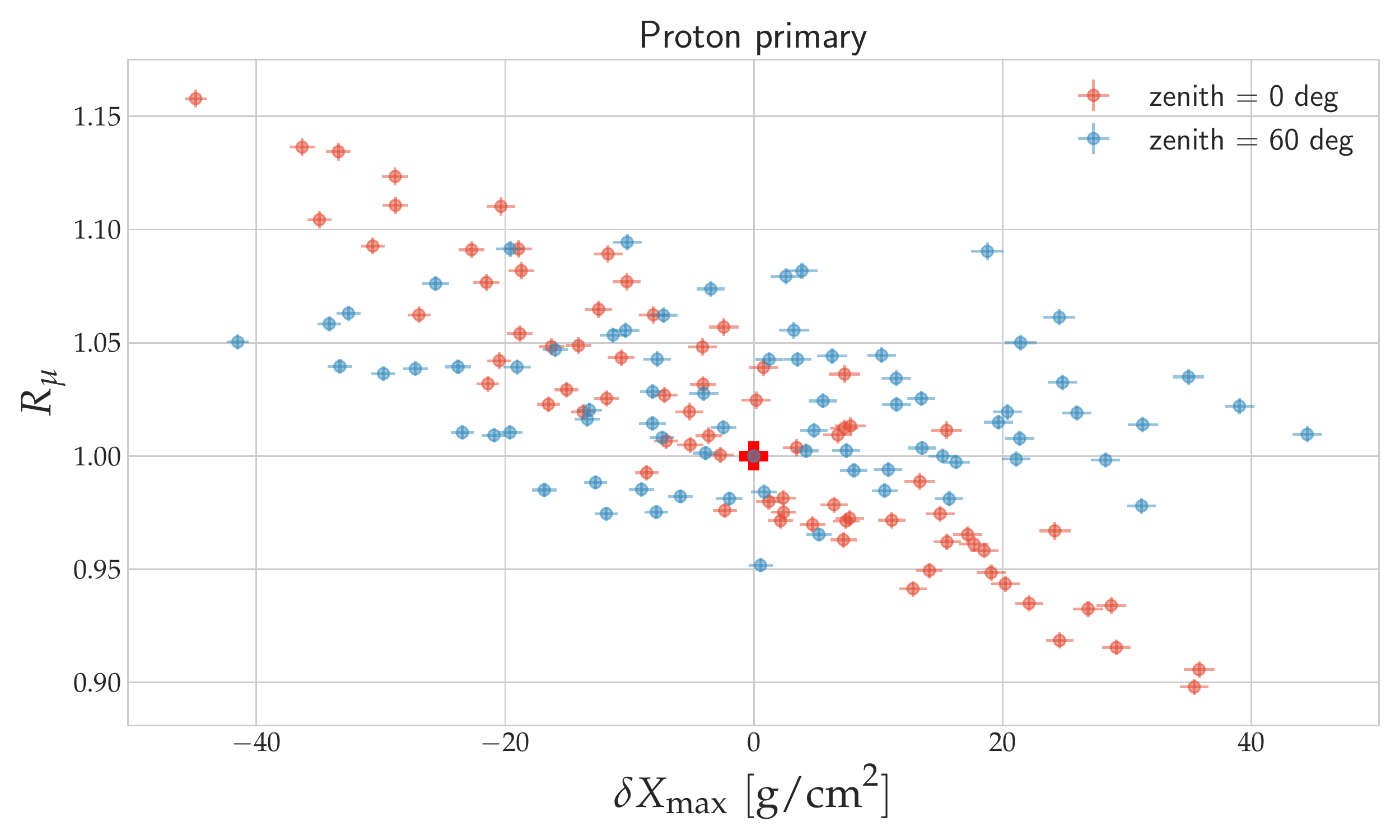}
\caption{Results of simulations in [$\delta X_{\mathrm{max}}$, $R_{\mu}$], proton primary.}
\label{basic_proton}       
\end{figure}

\begin{figure}[h]
\centering
\includegraphics[width=8cm,clip]{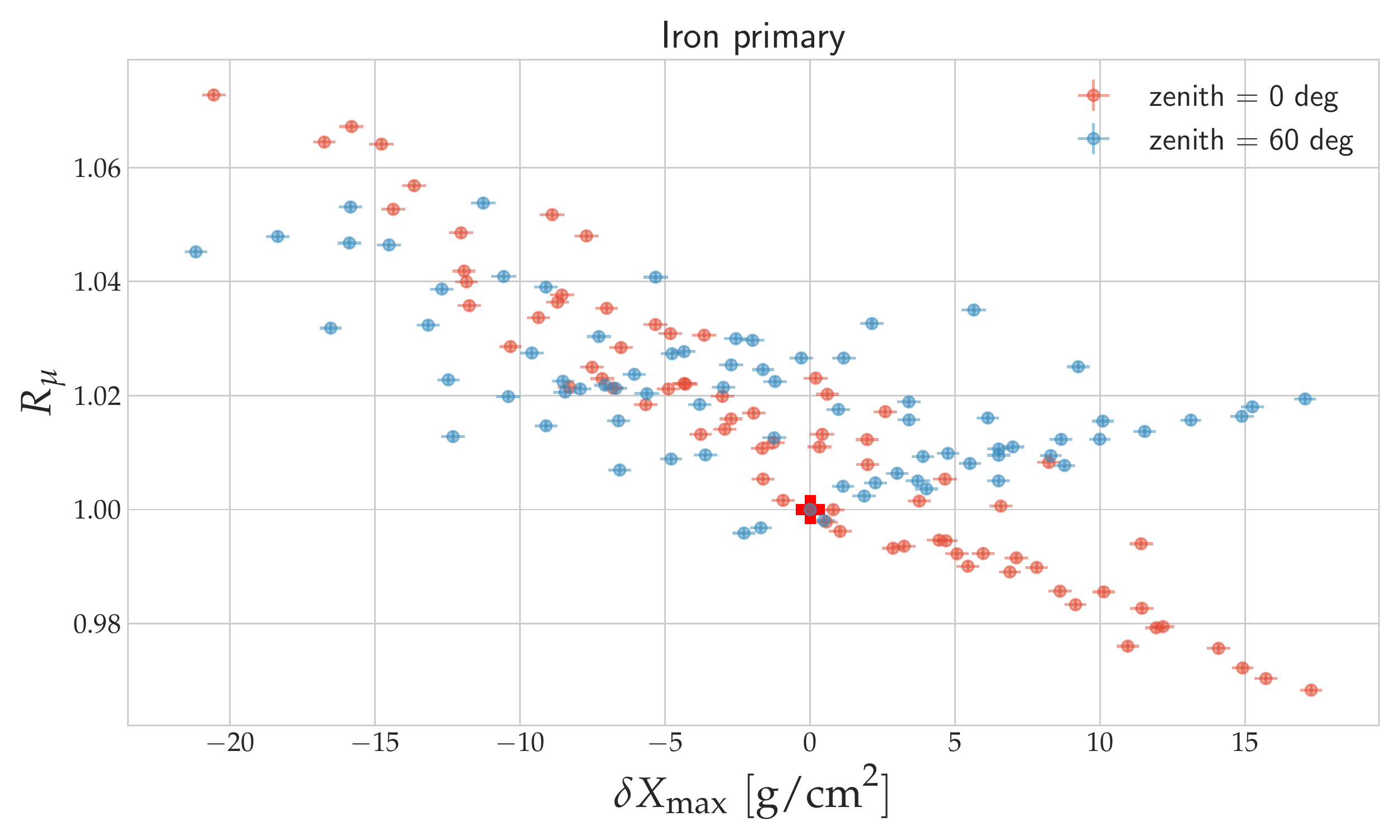}
\caption{Results of simulations in [$\delta X_{\mathrm{max}}$, $R_{\mu}$], iron primary.}
\label{basic_iron}       
\end{figure}

Because the exact values of \Xmax and $N_\mu^{1000}$ fluctuate from shower to shower, we consider the following four observables: the mean shift in \Xmax: $\delta X_{\mathrm{max}} = X_{\mathrm{max}} - X_{\mathrm{max}}(\mathrm{ref})$, its  standard deviation $\sigma(\Xmax)$, the muon rescaling: $R_{\mu} = N_{\mu}^{1000}/N_{\mu}^{1000}(\mathrm{ref})$ and the change in the standard deviation of the muon number $\sigma(N_\mu^{1000})/\sigma(N_\mu^{1000}(\mathrm{ref}))$, where the reference values are determined from the unmodified simulations in which $f^{\sigma}_{19} = f^{\mathrm{el}}_{19} = f^{\mathrm{mult}}_{19} = 1$. Note that while absolute \Xmax and $\sigma(\Xmax)$ values in g/cm$^2$ have a straightforward interpretation, the numbers for $N_\mu^{1000}$ are usually not measured directly and thus are best expressed as dimensionless ratio with respect to the reference value.

\begin{figure}[h]
\centering
\includegraphics[width=8cm,clip]{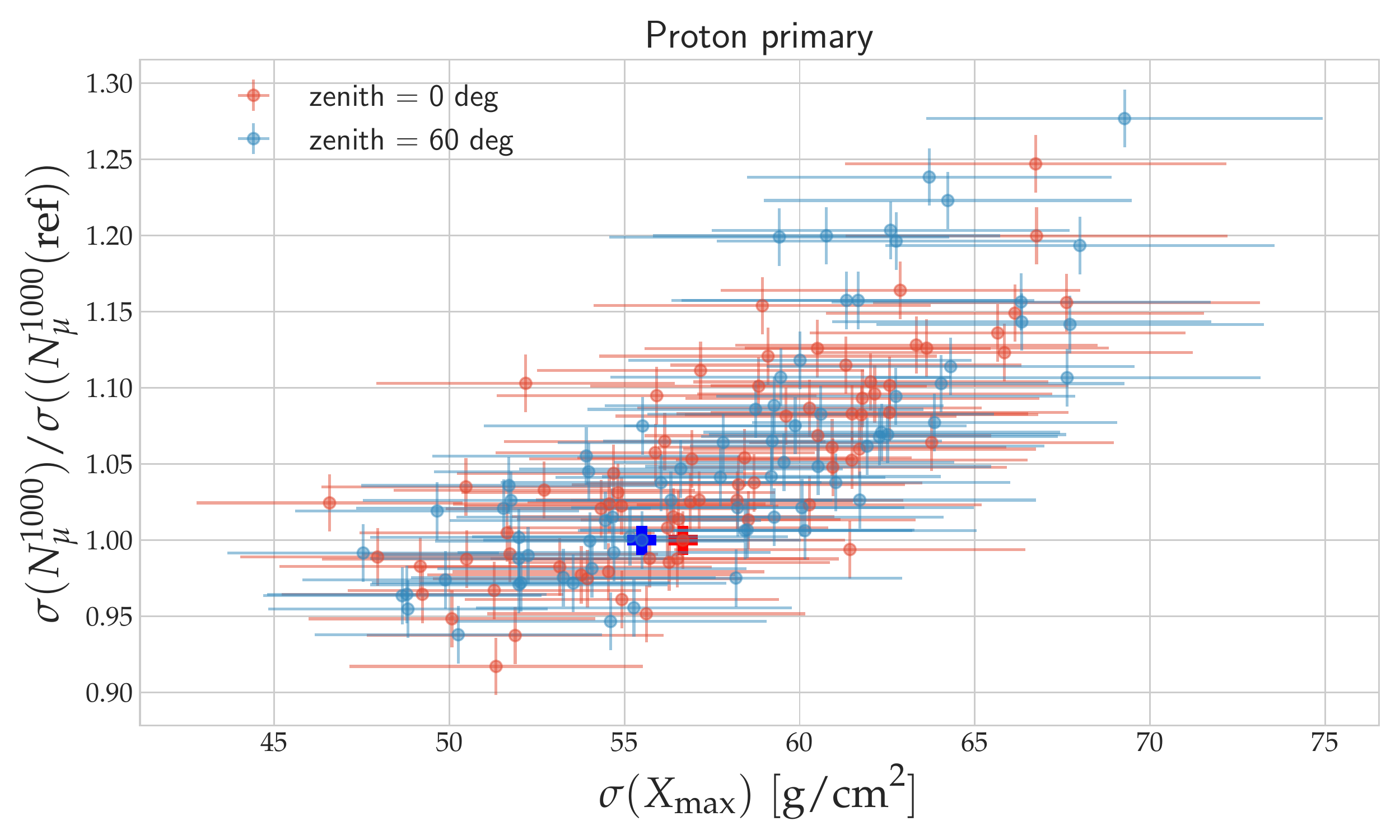}
\caption{Results of simulations in [$\sigma(\Xmax)$, $\sigma(N_\mu^{1000})/\sigma(N_\mu^{1000}(\mathrm{ref}))$], proton primary.}
\label{sigmas_proton}       
\end{figure}

\begin{figure}[h]
\centering
\includegraphics[width=8cm,clip]{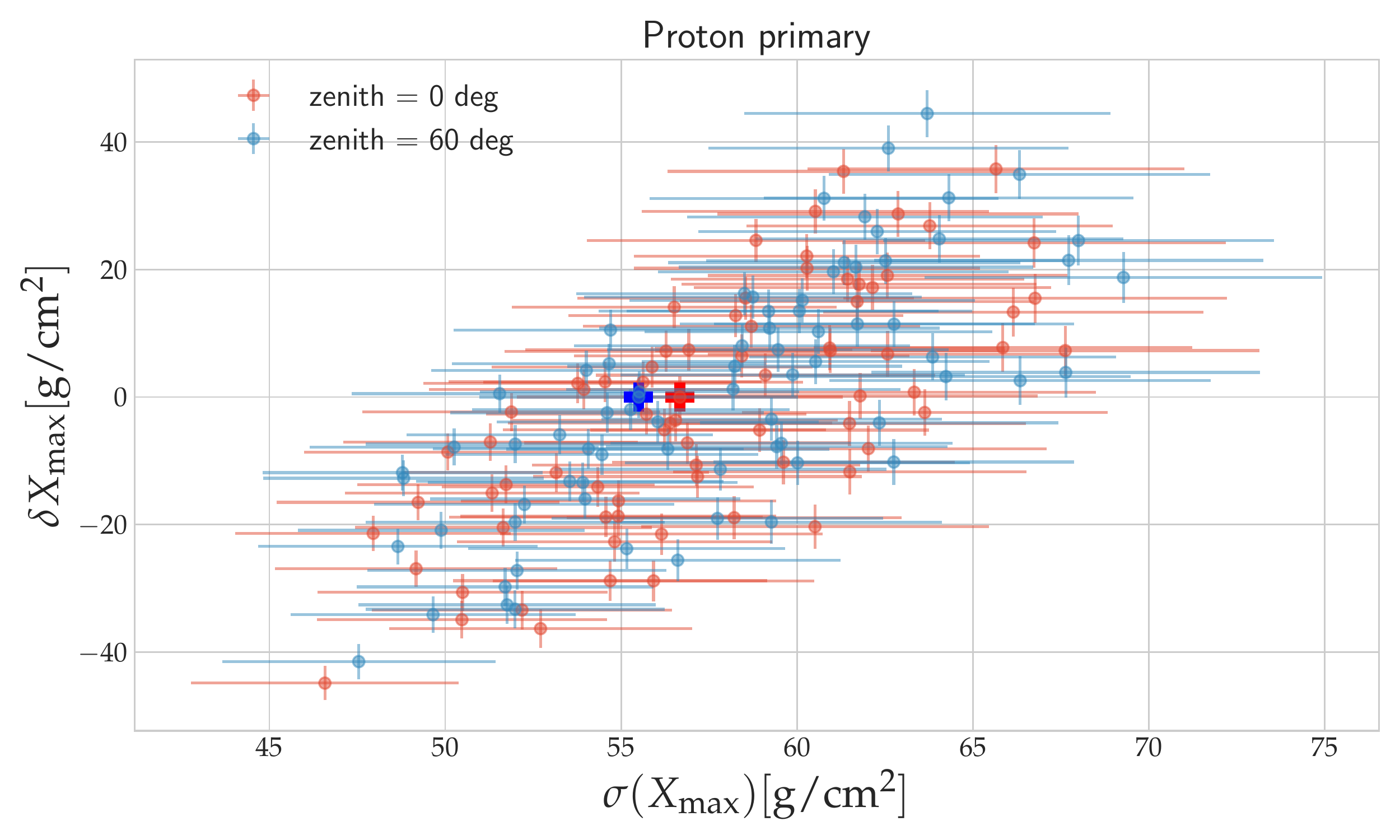}
\caption{Results of  simulations in [$\sigma(\Xmax)$, $\delta X_{\mathrm{max}}$], proton primary.}
\label{sigma_Xmax_vs_delta_Xmax_proton}       
\end{figure}

\begin{figure}[h]
\centering
\includegraphics[width=8cm,clip]{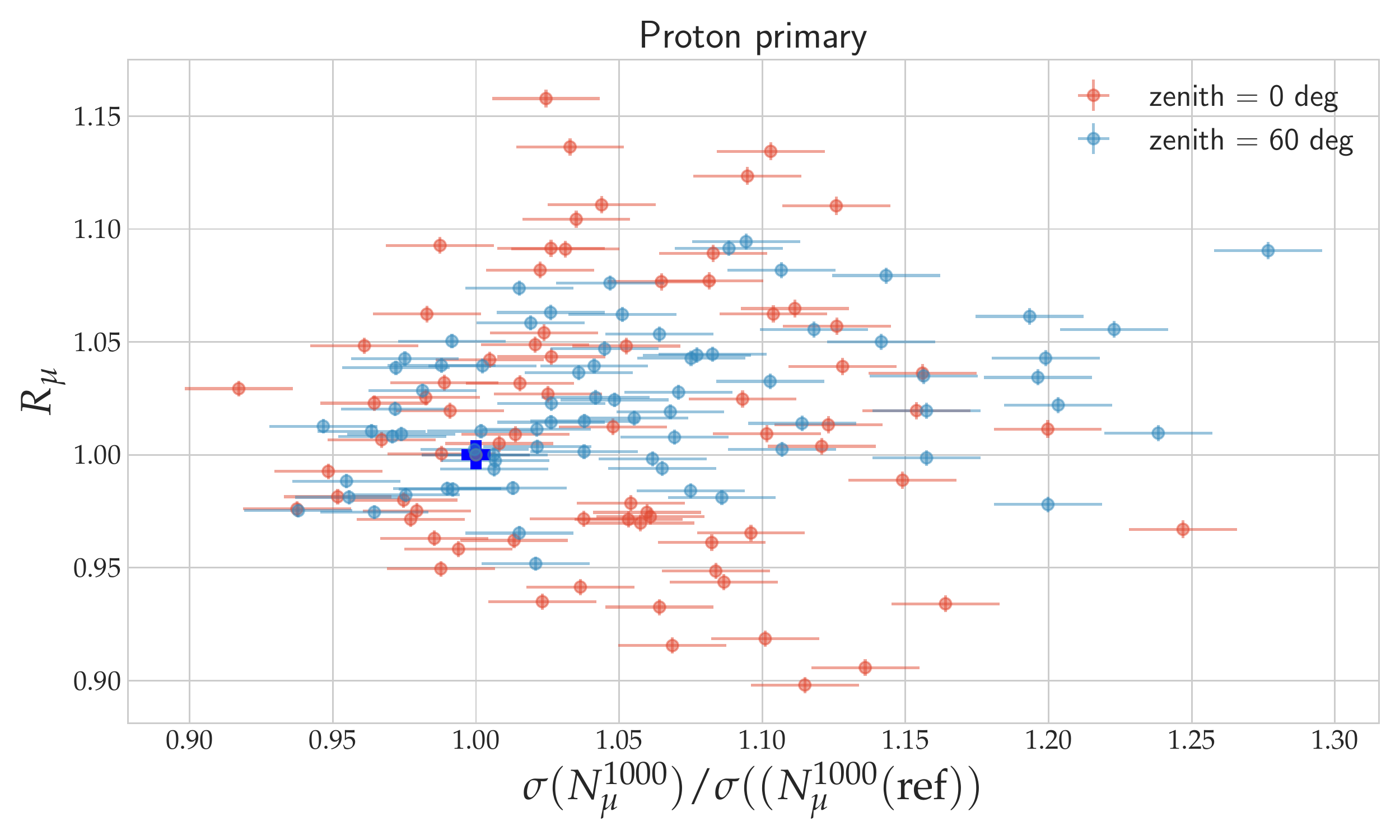}
\caption{Results of simulations in [$\sigma(N_\mu^{1000})/\sigma(N_\mu^{1000}(\mathrm{ref}))$, $R_{\mu}$], proton primary.}
\label{sigma_mu_Rmu_protons}       
\end{figure}

\begin{figure}[h]
\centering
\includegraphics[width=8cm,clip]{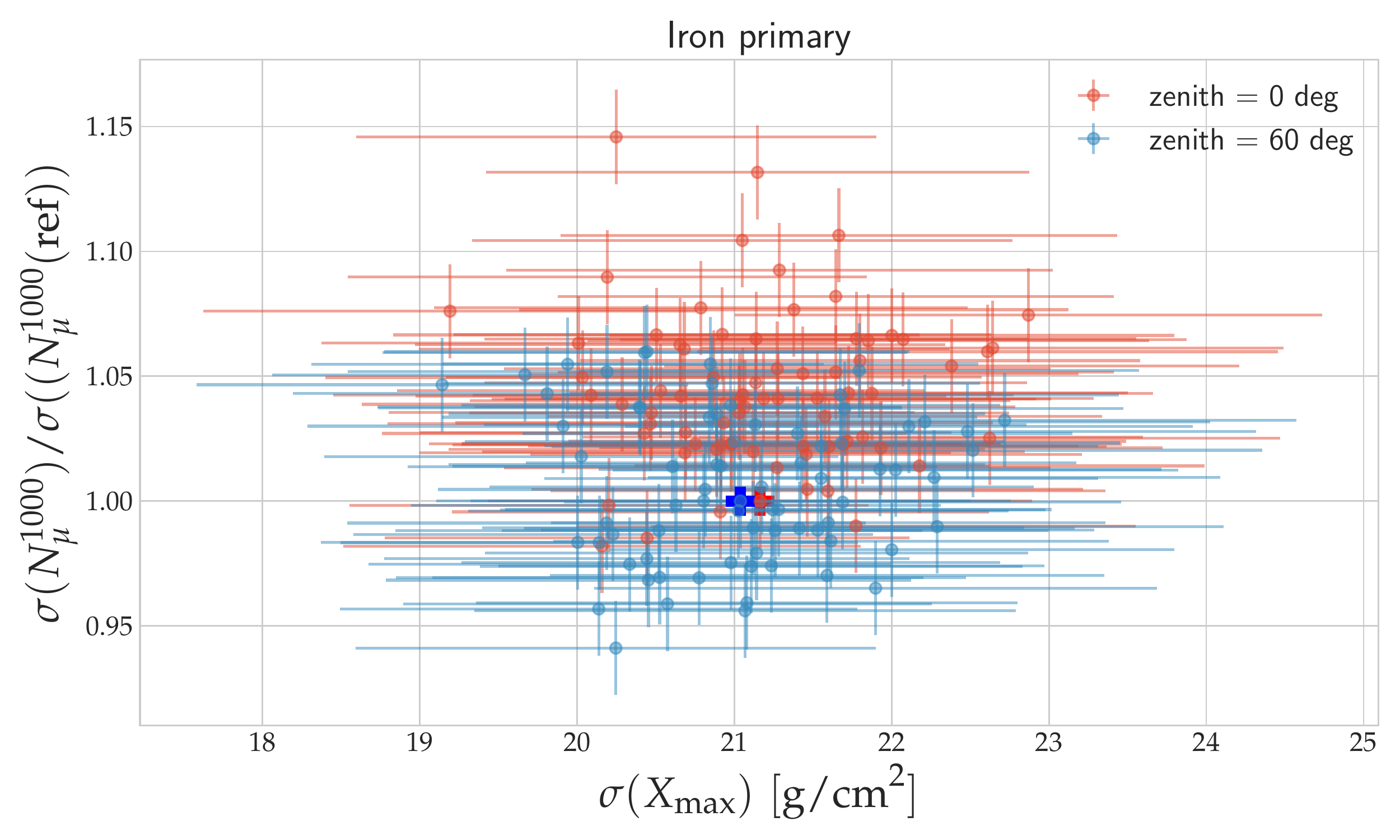}
\caption{Results of simulations in [$\sigma(\Xmax)$, $\sigma(N_\mu^{1000})/\sigma(N_\mu^{1000}(\mathrm{ref}))$], iron primary. Note the difference in scale compared to Fig. \ref{sigmas_proton}.}
\label{sigmas_iron}       
\end{figure}

\begin{figure}[h]
\centering
\includegraphics[width=8cm,clip]{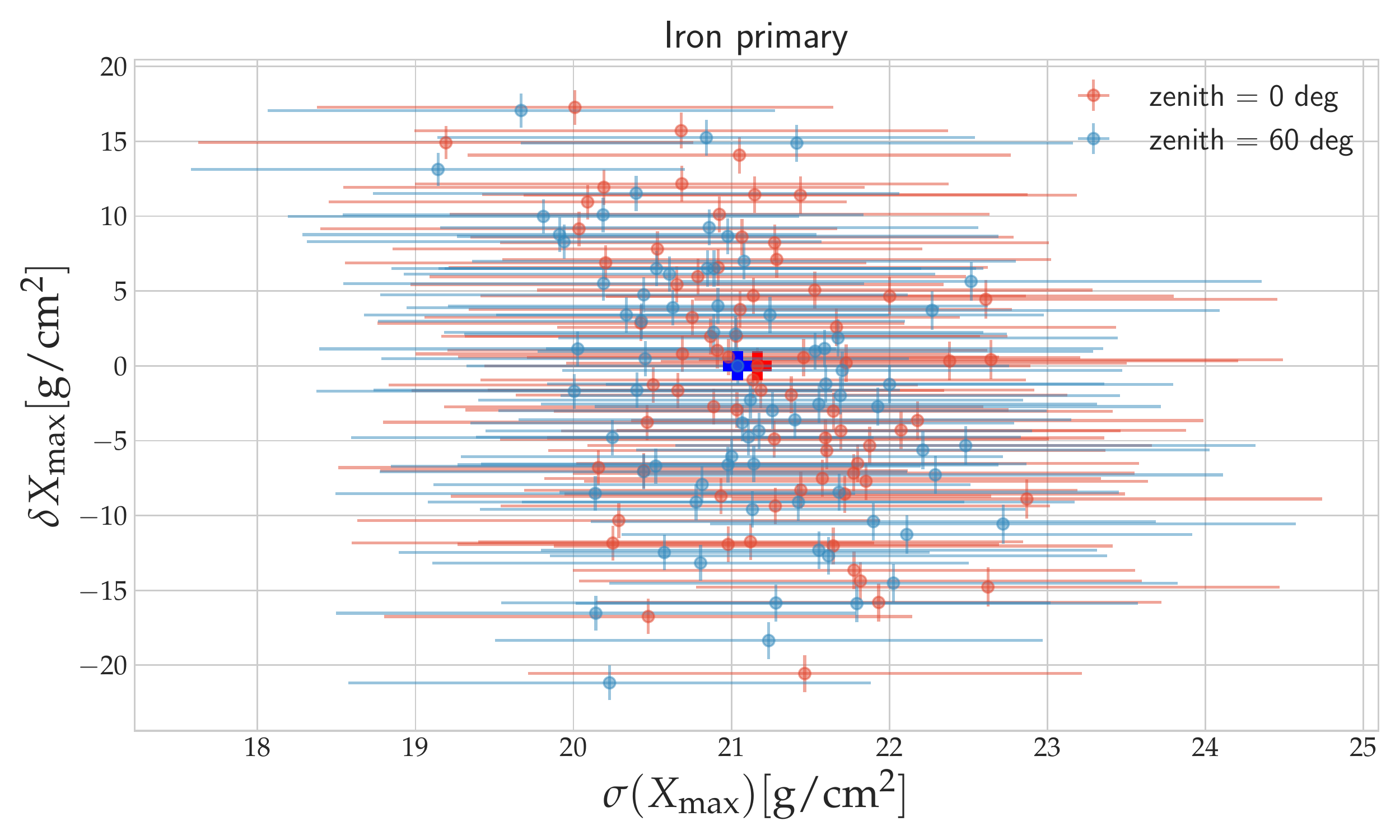}
\caption{Results of simulations in [$\sigma(\Xmax)$, $\delta X_{\mathrm{max}}$], iron primary.}
\label{sigma_Xmax_vs_delta_Xmax_iron}       
\end{figure}

\begin{figure}[h]
\centering
\includegraphics[width=8cm,clip]{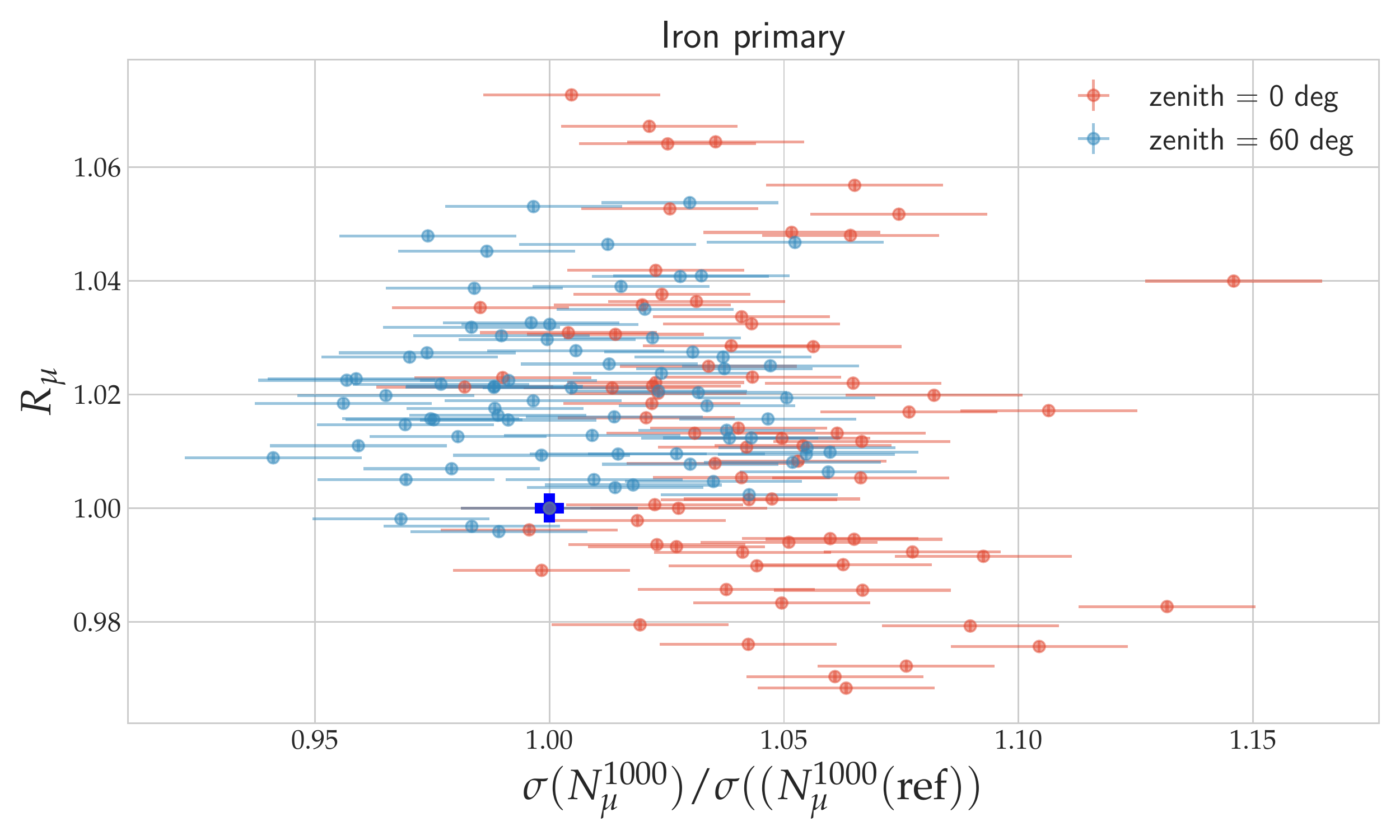}
\caption{Results of simulations in [$\sigma(N_\mu^{1000})/\sigma(N_\mu^{1000}(\mathrm{ref}))$, $R_{\mu}$], iron primary.}
\label{sigma_mu_Rmu_iron}       
\end{figure}

Each of Figs.~\ref{basic_proton}--\ref{sigma_mu_Rmu_iron} shows the results of all the simulations, with each point representing one of the possible combinations of [$f^{\sigma}_{19}, f^{\mathrm{el}}_{19}, f^{\mathrm{mult}}_{19}$] projected on a different plane in the [$\delta X_{\mathrm{max}}$,  $\sigma(\Xmax)$,  $R_{\mu}$, $\sigma(N_\mu^{1000})/\sigma(N_\mu^{1000}(\mathrm{ref}))$] space. In each of the plots, thick crosses indicate the reference values for unmodified simulations $f^{\sigma}_{19} = f^{\mathrm{el}}_{19} = f^{\mathrm{mult}}_{19} = 1$. These plots show both the extent of the possible changes of the observables within the range of the considered modifications and the correlations between different observables. Note that the ranges of the axes differ for proton and iron primaries; the effects of the modifications on iron primaries are universally smaller, as expected from the way the superposition model is applied to generate them. 

In \cite{Blazek:2021Cb} we showed the effect of each the modifications in detail, here we shall only qualitatively describe their direction in the phase space. Changes in cross-section move the points along the $\delta X_{\mathrm{max}}$ axis, as increasing the cross-section leads to lower $\delta X_{\mathrm{max}}$, i.e. shallower showers, as expected, with only a small effect on the number of muons produced. Changes in multiplicity and elasticity result in anti-correlated shifts in $X_{\mathrm{max}}$ and $R_{\mu}$ (along the anti-diagonal in out plots), but for multiplicity, the change in $X_{\mathrm{max}}$ is, for the same change in $R_{\mu}$, much smaller than for elasticity.

\begin{figure*}
\centering
\vspace*{5cm}       
\includegraphics[width=14cm,clip]{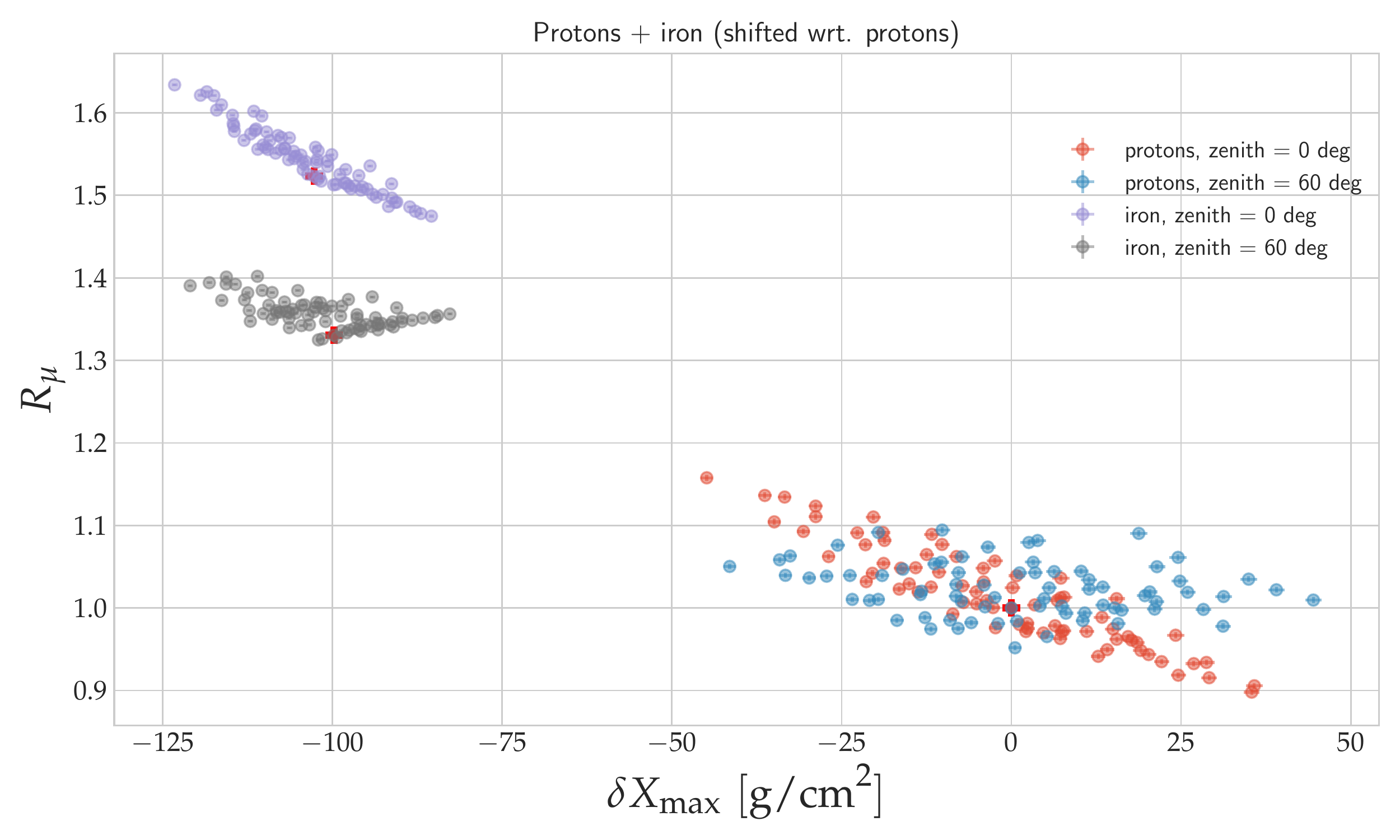}
\caption{Mean muon rescaling $\Rmu$ and mean shift of $\Xmax$ for proton and iron primaries shown in both cases with respect to unmodified simulations for proton. Results are shown for two zenith angles, in each case the reference is taken at the corresponding zenith angle; each point corresponds to a particular combination of $f_{19}^{\mathrm{sigma}}$, $f_{19}^{\mathrm{el}}$ and $f_{19}^{\mathrm{mult}}$.}
\label{scatter_combined}       
\end{figure*}

Figs.~\ref{basic_proton} and \ref{basic_iron} show that for vertical showers, there is a strong correlation where an increase in the number of muons implies shallower showers. This is particularly interesting in view of the recent results from the Pierre Auger Observatory \cite{kuba} which indicate that the data is best described if the number of muons and \Xmax in simulations are increased simultaneously, which is very difficult to achieve with any combination of hereby considered modifications. Similarly interesting is the observation that for inclined showers the effect of the modification on the number of muons is generally smaller.
Figs.~\ref{sigmas_proton}--\ref{sigma_mu_Rmu_iron} show that the effect of modifications on the variances of the variables is far more pronounced for proton than for iron - in particular $\sigma(\Xmax)$ for iron is very difficult to change by any modifications. For protons, the changes in \Xmax itself, $\sigma(\Xmax)$ and $\sigma(N_\mu^{1000})$ are well correlated, showing again (similarly to the case of the correlation between $R_{\mu}$ and \Xmax discussed above) that even making parallel modifications in several significantly different ways still enables the coverage of only a restricted region of the space of possible values of observables. Such correlations may have deep implications when compared with experimental data.

Fig. \ref{scatter_combined} presents the same data as Figs.~\ref{basic_proton} and \ref{basic_iron} but in a way that allows for additional insight. The values for muon rescaling $\Rmu$ and mean shift of $\Xmax$ for proton and iron primaries for all the applicable combinations of $f_{19}^{\mathrm{sigma}}$, $f_{19}^{\mathrm{el}}$ and $f_{19}^{\mathrm{mult}}$ and two zenith angles are shown with respect to unmodified simulations \emph{for proton} at the corresponding zenith angle. From this plot it is immediately clear that the effects of the modifications on iron showers are significantly smaller as expected. It also shows that even for unmodified simulations (highlighted by thick crosses), the difference in $\Rmu$ between iron and proton primaries is much more pronounced for vertical than for inclined showers; the same is then true for the influence of the modifications on $\Rmu$. Secondly, one can see that for vertical showers, the points are concentrated along a line that almost, but not quite follows the line between proton and iron. Such a trend is expected, as increasing multiplicity and cross-section makes proton act more “iron-like”, but at the same time it is clear that there is more to the difference between proton and iron showers than simply the difference in the three macroscopic parameters. This is even more pronounced for inclined showers where the proton points are arranged along an almost horizontal axis. This fact is actually encouraging for the outlooks for determining the properties of hadronic interactions from UHECR data because it shows that there is not a complete degeneracy between the – both unknown – chemical composition and properties of hadronic interactions of the cosmic rays, at least not for the three basic properties of cross-section, multiplicity and elasticity.

\section{Conclusions and Outlook}

We have shown that ad-hoc modifications of macroscopic properties of a model of hadronic interactions, carried out within the currently applicable experimental constraints, allow for substantial changes in air-shower observables, however only for certain combinations of these observables. In particular we have shown that achieving an increase in the number of muons on the ground  at 1000 meters from the shower axis simultaneously with an increase in the atmospheric depth of the shower maximum is challenging.  We have demonstrated that such ad-hoc changes can be implemented in a consistent way for proton and nuclear primaries without the need to introduce additional parameters, if the superposition model of nuclear interactions is adopted. This opens the door to further studies with mixed primary beams that would provide more realistic description of the real cosmic-ray flux, more suitable for direct comparison with experiments.

In further work, we plan to study the effects of the modifications on the signal at different distances from the shower core and at different energy thresholds for the detection of secondary particles at the ground as well as on muon production depth, frequency of anomalous showers, the correlation between \Xmax and the ground signal and further observables. We can also determine which observables are the most sensitive to the individual parameters of the hadronic interactions, thus helping to guide the design of future cosmic-ray observatories. 

\section*{Acknowledgements}
\begin{acknowledgement}
 This work is funded by the Czech Science Foundation under the project $\rm GACR 21-02226M$.
\end{acknowledgement}

\hfill

\bibliography{bibtex} 

\end{document}